\documentclass[12pt]{iopart}

\usepackage{graphicx}
\usepackage{iopams}  
\usepackage{color}

\begin{document}

\title[Path integral for Darcy flow through random porous media]{Pressure statistics from the path integral for Darcy flow through random porous media}

\author{Marise J E Westbroek$^{1,2}$, Gil-Arnaud Coche$^3$, Peter R King$^1$, and Dimitri D Vvedensky$^2$}
\address{$^1$ Department of Earth Science and Engineering, Imperial College London, London SW7 2BP, United Kingdom}
\address{$^2$ The Blackett Laboratory, Imperial College London, London SW7 2AZ, United Kingdom}
\address{$^3$ Accuracy, 41 Rue de Villiers, 92200 Neuilly-sur-Seine, France}

\begin{abstract}
The path integral for classical statistical dynamics is used to determine the properties of one-dimensional Darcy flow through a porous medium with a correlated stochastic permeability for several spatial correlation lengths.  Pressure statistics are obtained from the numerical evaluation of the path integral by using the Markov chain Monte Carlo method.  Comparisons between these pressure distributions and those calculated from the classic finite-volume method for the corresponding stochastic differential equation show excellent agreement for Dirichlet and Neumann boundary conditions.  The evaluation of the variance of the pressure based on a continuum description of the medium provides an estimate of the effects of discretization. Log-normal and Gaussian fits to the pressure distributions as a function of position within the porous medium are discussed in relation to the spatial extent of the correlations of the permeability fluctuations. 
\end{abstract}

%

\section{Introduction}

Flow in porous media arises in many branches of science, engineering, and industry, including geologic CO$_2$ sequestration \cite{juanes12}, enhanced oil recovery \cite{orr84}, and water infiltration into soil \cite{juanes08}.  Examples involving porous biological tissue \cite{khaled03} include the flow of oxygen through lungs \cite{miguel12}, the transport of cerebrospinal fluid through the brain \cite{penn07} and hemodynamics through blood vessels \cite{zunino10}. Artificial porous media in widespread industrial use include building materials \cite{hu13} and catalysts \cite{perego13}. The main computational concern in many applications is based around how the pore structure affects the flow characteristics through the medium. The path integral methodology we describe here is appropriate for slow, single-phase flow, in a porous material with a correlated random permeability.  In particular, we consider data that are typical for flow in hydrocarbon reservoirs and groundwater aquifers.

The permeability of a reservoir rock, which describes its resistance to fluid flow, relates the local flow rate to the local pressure gradient.  Permeability data are available at only a limited number of points in the rock.  For hydrocarbon and groundwater flow applications, the presumed properties of the rock away from those locations can sometimes be inferred from analogue outcrops.  The most common way of obtaining a quantitative description of the permeability, however, is through the calibration of a model to observed data.  The aim of any reservoir engineering model is to generate simulations of the pressure, flow rates and fluid saturations in terms of probability distributions in a random porous medium. 

In this paper, we focus on the simplest case of single-phase, steady-state incompressible flow.  There are two basic methods to simulate such flow through a porous medium  \cite{renard}.  One is to solve the Navier-Stokes equations at the pore level \cite{aramideh18}.  This method is computationally too intensive for our purposes, because our goal is to simulate the flow in a macroscopic volume that spans many pores. The alternative is to invoke Darcy's law \cite{darcy56,whitaker86}, an empirical linear relation between the average velocity of the fluid $\bi{q}$, the pressure gradient $\bi{\nabla} p$, and the effective permeability $K$ of the flow through a porous medium:
\begin{equation}
\label{Darcy}
\bi{q}=-K \nabla p\, .
\end{equation}
Here, $K(\bi{x})=k(\bi{x})/\mu$, where $k(\bi{x})$ is the effective permeability of the medium and $\mu$ is the viscosity of the fluid.  Darcy's law is valid on a ``mesoscopic'' scale, large compared to the pore scale, but small on the scale of the macroscopic system.  To solve Darcy's law, we must have a realistic representation of $K(\bi{x})$.  We use a log-normal model for the permeability \cite{law,freeze}. For an incompressible fluid, $\bi{\nabla}\cdot \bi{q}=0$, which, together with Darcy's law (\ref{Darcy}), yields:
\begin{equation}
\label{gradzero}
\bi{\nabla}\cdot(K(\bi{x})\bi{\nabla} p(\bi{x}))=0.
\end{equation} 

The established way of calculating pressure statistics is by generating discrete realizations of the porous medium and solving for the pressure field using (\ref{gradzero}).  This way of discretizing and evaluating a differential equation is known as the finite-volume method (FVM) \cite{schafer}.  Here, we propose an alternative approach, based on the path integral formalism, which does not rely on explicit permeability realizations.  Instead, we work directly with the probability distribution of the log-permeability field.  Subject to Darcy's law, we generate pressure realizations.  Both methods require a large number of independent realizations for a reliable error estimate. Casting the problem in path integral form has the advantage of access to existing analytic techniques, notably the renormalization group and applications of perturbation theory \cite{amit84, zinn-justin06}, both of which have been applied to Darcy flow \cite{tanksley94,hanasoge17}, as well as an abundance of numerical methods for the evaluation of path integrals \cite{morningstar,westbroek18a}, which is an altogether new approach to this problem \cite{westbroek18b}.

The organization of our paper is as follows.  Section~\ref{sec2} explains how a random permeability is described within Darcy's law for a one-dimensional system, including the generation of correlated log-normal statistics for the permeability.  There are two ways of solving Darcy's law for a one-dimensional system:~the direct solution of Darcy's law (\ref{gradzero}) with a random permeability, and the evaluation of the path integral, which is the approach taken here.  The direct solution of Darcy's law by the finite-volume method is described in Sec.~\ref{sec3}, and the formulation and numerical solution of the path integral for Darcy's equation with the Markov chain  Monte Carlo method are developed in Secs.~\ref{sec4} and \ref{sec5}.  The pressure statistics obtained with the two methods are compared for Neumann and Dirichlet boundary conditions in Secs.~\ref{sec6} and \ref{sec7}, with an emphasis on the role of the correlation length in determining the pressure distributions.  Section~\ref{sec8} summarizes our results and provides a brief description of future work.  The appendices contain derivations for our statistical analysis.

\section{Darcy's Law with Stochastic Permeability}
\label{sec2}

\subsection{Solutions with Dirichlet and Neumann Boundary Conditions}
\label{sec2.1}


\begin{figure}[t]
\centering
\includegraphics[width=0.45\textwidth]{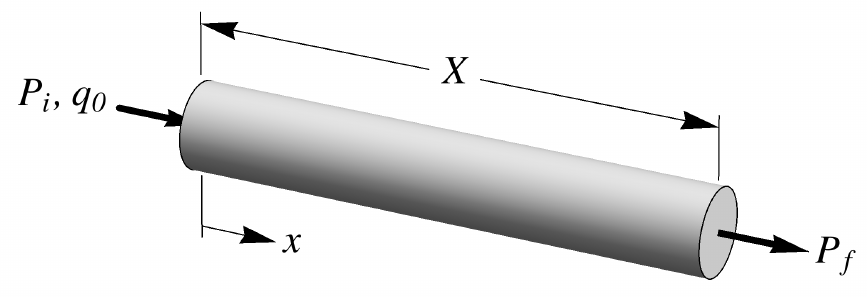}
\caption{Schematic depiction of one-dimensional flow through a permeable medium of length $X$. The term ``one-dimensional'' refers to the number of spatial coordinates $x$ required to describe the flow. That is, flow is along the $x$-direction, with no flow in the lateral directions.  For Dirichlet boundary conditions, the initial and final pressures $P_i$ and $P_f$ are specified at the entry and exit points of the medium, respectively, while for Neumann boundary conditions, $P_i$ and the flow $q_0$ are specified at the entry to the medium, leaving the final pressure determined by the realization of the permeability fluctuations.}
\label{fig1}
\end{figure}


We consider viscous flow along the $x$-direction through a one-dimensional permeable ``rock'' of length $X$. The one-dimensional form of Darcy's law (\ref{Darcy}) reads:
\begin{equation}
\label{darcy}
q(x)=-K(x)\frac{d p(x)}{dx}\, .
\end{equation}
In one dimension, incompressible flow reduces to constant flow, $q(x)=q_0$, so integrating (\ref{darcy}) subject to the initial condition $p(0)=P_i$, yields\begin{equation}
\label{DarcyNBC}
p(x)=P_i-q_0 R(x), 
\end{equation}
where
\begin{equation}
\label{eq:R}
R(x)=\int_0^x\frac{dx^\prime}{K(x^\prime)}
\end{equation}
is a resistance to flow. By setting $x=X$ in (\ref{DarcyNBC}), we obtain a relation between the initial flow $q_0$ and the final pressure $p(X)=P_f$:
\begin{equation}
\label{pa}
q_0=-\frac{P_f-P_i}{R(X)}\, .
\end{equation}
To solve for $p(x)$, one additional variable must be specified.  Imposing $q_0$, the derivative of $p(x)$ at the boundary, defines a Neumann boundary condition (NBC).  In oil-field terms, fixed $q_0$ corresponds to a constant injection rate. The specification of $P_f$, the pressure at the exit of the well (Fig.~\ref{fig1}), is the condition of constant production, and yields a Dirichlet boundary condition (DBC).

\subsection{Correlated Permeability Fluctuations}
\label{sec2.2}


\begin{figure}
\centering
\hskip2cm\includegraphics[width=0.8\textwidth]{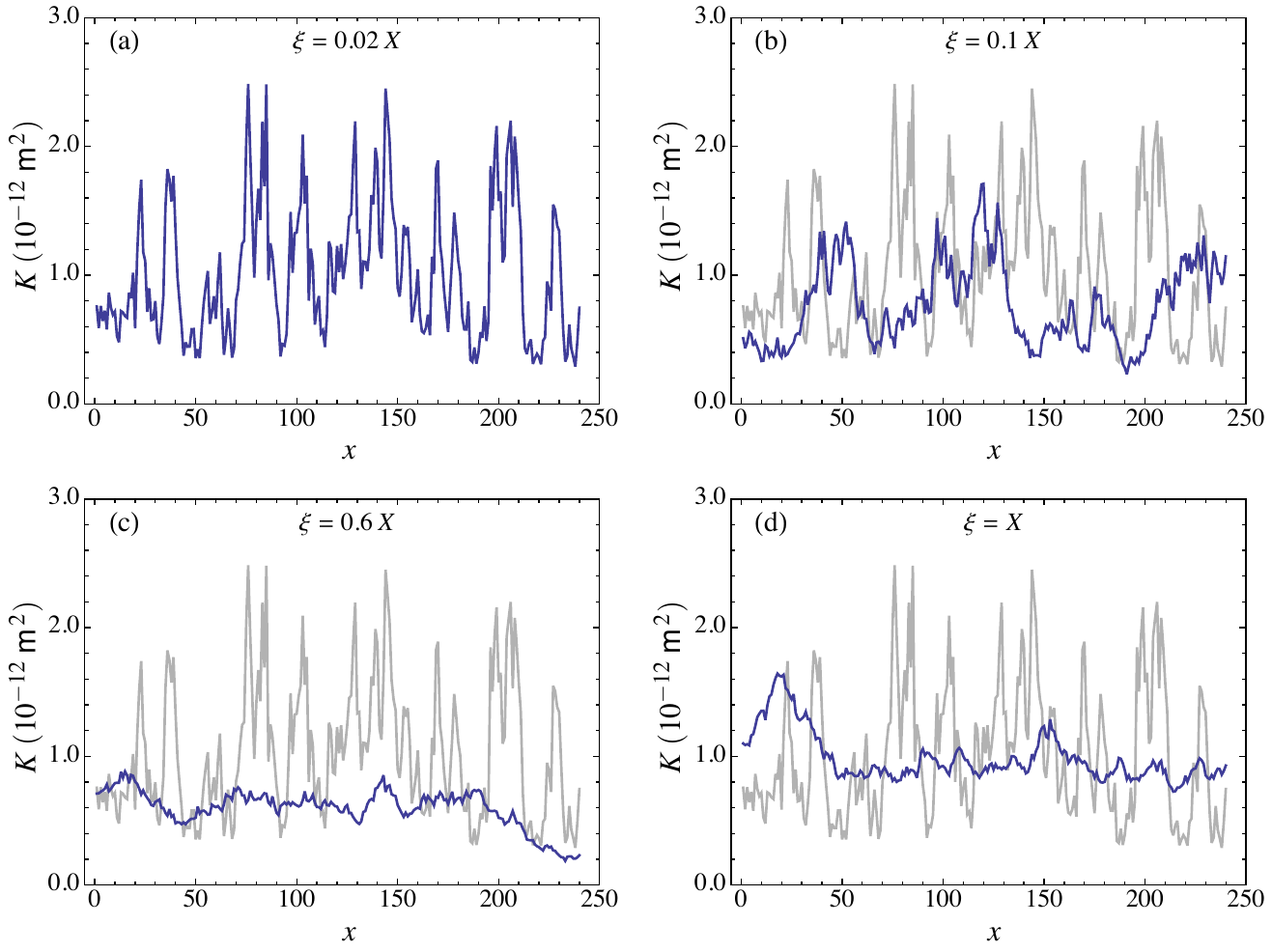}
\caption{Realizations of the effective permeability $K(x)$ for the indicated correlation lengths obtained from a stationary Ornstein--Uhlenbeck process with the correlation function (\ref{eq:covL}). For $\xi=0.02X$ (a), permeability correlations are barely visible. For $\xi=X$ (d), such correlations are apparent in the small site-to-site variations. In (b), (c), and (d), the profile in (a) shown in gray to emphasize the contrasting scale and magnitude of permeability fluctuations as a function of the correlation length. Simulations were carried out on systems of $N_x=240$ sites with $\Delta x=0.5$~m.}
\label{fig2}
\end{figure}


With the log permeability of the rock denoted by $L(x)\equiv\log K(x)$, the Gaussian random variable $L(x)$ is modelled as a stationary Ornstein--Uhlenbeck process \cite{sahimi93}. Such a process is fully characterized by its mean (which we set to zero) and covariance function
\begin{equation}
\label{eq:covL}
\mathrm{Cov}(L(x),L(y))=\sigma^2 e^{-|x-y|/\xi}.
\end{equation}
The choice to set the mean of the log-permeability to zero implies that the permeability has geometric mean one. The correlation length $\xi$ describes the typical length scale over which the permeabilities take comparable values;~we have set $\sigma=0.5$ for all simulations presented here. More information about the Ornstein-Uhlenbeck process, including a derivation of (\ref{eq:covL}) may be found in \ref{secA1}. 

The effect of the correlation length is illustrated in Fig.~\ref{fig2}.  For a correlation length of a few lattice spacings (Fig.~\ref{fig2}(a)), the permeability is effectively random on neighboring sites, and so shows substantial fluctuations over small distances.  As the correlation length increases to the system size (Fig.~\ref{fig2}(b,c,d)), the site-to-site variations of the permeability are significantly diminished, as is the magnitude of the fluctuations through the system, though appreciable variations can still occur over larger distances.

\subsection{Pressure Profiles}
\label{sec2.3}

\begin{figure}[t]
\centering
\hskip2cm\includegraphics[width=0.8\textwidth]{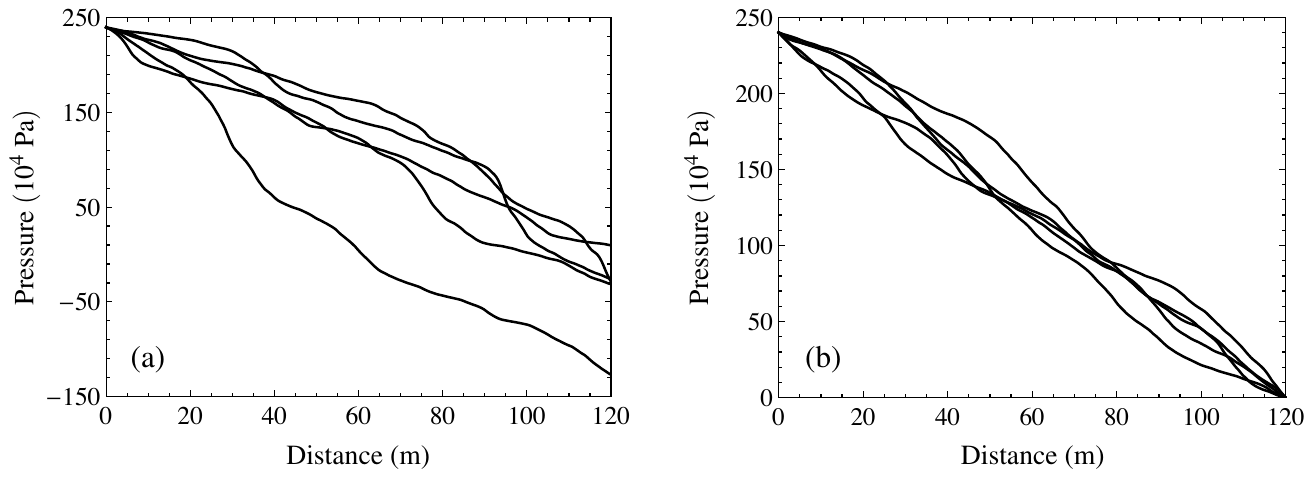}
\caption{Pressure profiles with (a) Neumann and (b) Dirichlet boundary conditions for $\xi=0.1 X$. Simulations were carried out on systems of $N_x=120$ sites with $\Delta x=1$. The pressure gauge invariance of Darcy's equation means that only pressure differences are meaningful;~negative pressures result from a particular choice of a zero of pressure.}
\label{fig3}
\end{figure}

Examples of pressure profiles in systems with Neumann and Dirichlet boundary conditions are shown in Fig.~\ref{fig3}. In both cases the pressure decreases monotonically due to the resistance of the permeable medium.  The difference is that, for DBC, the pressure is fixed at both ends of the system while, for NBC, the pressure is fixed only at entry (along with the injection rate), so the exit pressure is a free variable. In particular, the distribution of pressures broadens along the system for NBC, but, for DBC, broadens initially, then narrows as the effect of the exit boundary takes hold.  Pressure distributions will be analyzed in Secs.~\ref{sec6} and \ref{sec7}.

\section{Finite-Volume Method}
\label{sec3}


The statistics of pressure distributions in a one-dimensional permeable medium are given in terms of the permeability distribution by the solution (\ref{DarcyNBC}) and (\ref{eq:R}) to Darcy's equation (\ref{darcy}).  In two and three dimensions, the corresponding solutions of (\ref{gradzero}) are divergence-free vector fields expressed as $K(\bi{x})\bi{\nabla}p(\bi{x})=\bi{\nabla}\times\bi{A}$, where $\bi{A}$ is a vector potential.  These do not yield the simple analytic forms of (\ref{DarcyNBC}) and (\ref{eq:R}), so we must look to numerical solutions of (\ref{gradzero}) for each realization of the permeability, then average the results.  In this section, we describe the numerical method for accomplishing this, which we will then apply to the one-dimensional Darcy equation (\ref{darcy}).

The finite volume method (FVM) is a discretization technique for partial differential equations, especially those associated with conservation laws, such as (\ref{gradzero}). The FVM uses a volume integral formulation based on a partition of the system into volumes to discretize the equations by representing their solution as a set of algebraic equations for quantities defined within each volume.  Finite volume methods are based on applying conservation principles over each of the small volumes in the partition, so global conservation is ensured. The FVM is an established technique, especially for computational fluid dynamics, but we provide a brief description here for completeness.

\begin{figure}
\centering
\hskip2cm\includegraphics[width=0.6\columnwidth]{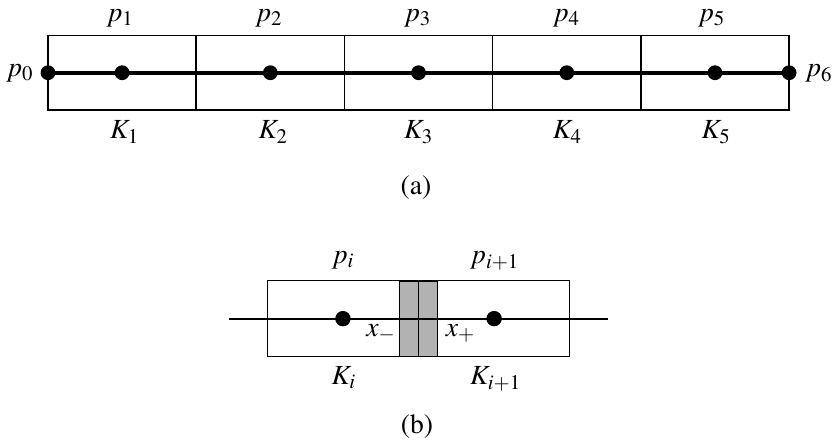}
\caption{(a) Schematic depiction of a one-dimensional lattice partitioned into 5 cells with $N_x+2=7$ sites according to the finite-volume method. The permeabilities $K_i$, for $i=1,2,3,4,5$, are considered constant within each cell, and the pressures at the cell boundaries are determined from (\ref{eq8}). (b) Section of the one-dimensional lattice with the integration region (shown shaded) used to derive the equation for the pressure.}
\label{fig4}
\end{figure}

Simulations for the FVM and the path integral are carried out on a lattice with $N_x+2$ sites (Fig.~\ref{fig4}(a)).  Solving (\ref{gradzero}) requires a permeability field $K(x)$.  The log-permeability $L(x)$ is generated first, after which the relation $K(x)=e^{L(x)}$ is used to obtain the permeabilities. 
There are various ways of generating $L(x)$.  A simple, but slow, method is the Cholesky decomposition of the covariance matrix for $L$ \cite{NR}, which takes $O(N_x^3)$ floating-point operations (flops) \cite{NR}. Another method, which is faster, but more involved, employs a fast Fourier transform (FFT) \cite{dietrich93,dietrich97}. The computational cost of the FFT method is $O(N_x \log N_x)$ flops.

Once the permeability field has been determined, we can derive an equation for the pressure by integrating (\ref{gradzero}) over a small region near the boundary between the $i$th and $(i+1)$st cells (Fig.~\ref{fig4}(b)):
\begin{eqnarray}
\int_{x_-}^{x_+}{d\over dx}\biggl(K{dp\over dx}\biggr)\,dx&=\biggl(K{dp\over dx}\biggr)\bigg|_{x_-}-\biggl(K{dp\over dx}\biggr)\bigg|_{x_+}\nonumber\\
&=K_i(p-p_i)-K_{i+1}(p_{i+1}-p)\, ,
\label{eq8}
\end{eqnarray}
where $p$ is the pressure at the boundary between the two cells. As there are no sources, the upper (resp.~lower) limit of integration can be extended to the right (resp.~left) edge of the system without affecting the value of the integral.  Hence, the two terms on the right-hand side must each be equal to the same constant: 
\begin{equation}
q_0=K_i(p-p_i)=K_{i+1}(p_{i+1}-p)\, .
\end{equation}
Elimination of the boundary pressure $p$ yields:
\begin{equation}
\label{harmonicmean}
q_0=\frac{K_i K_{i+1}}{K_i+K_{i+1}}(p_{i+1}-p_i)\equiv t_{i,i+1}(p_{i+1}-p_i)\, ,
\end{equation}
where we have defined the transmissibility $t_{i,i+1}$ as the harmonic mean of $K_i$ and $K_{i+1}$. For DBC, we set $p_0=P_i$ and $p_{N_x+1}=P_f$ while, for NBC, where $q_0$ is fixed, we only set $p_0=P_i$.  In each case, the solution of (\ref{gradzero}) is replaced by the solution of a set of linear equations of the form ${\sf T}_{D/N}\bi{P}=\bi{B}_{D/N}$,  where the subscript indicates Dirichlet ($D$) or Neumann ($N$) boundary conditions. The transmissibility matrices ${\sf T}_{D/N}$ are
\begin{equation}
\fl
{\sf T}_D=\left(
\begin{array}{ccccc}
t_{0,1}+t_{1,2}& -t_{1,2}&\cdots & 0 & 0\\
\noalign{\vskip6pt}
-t_{1,2}& t_{1,2}+t_{2,3}&\cdots & 0 & 0\\
\vdots&\vdots&\ddots&\vdots&\vdots\\
0& 0 &\cdots & t_{N_x-2,N_x-1}+t_{N_x-1,N_x} & -t_{N_X-1,N_x}\\
\noalign{\vskip6pt}
0 & 0 &\cdots & -t_{N_x-1,N_x} & t_{N_x-1,N_x}+t_{N_x,N_x+1}
\end{array}\right)\, ,
\end{equation}
for Dirichlet boundary conditions,
\begin{equation}
{\sf T}_N=\left(
\begin{array}{cccccc}
\hskip3ptt_{0,1} & 0 & \cdots & 0 & 0 & 0 \\
\noalign{\vskip6pt}
\hskip-6pt -t_{1,2} & t_{1,2} & \cdots & 0 & 0 & 0 \\
\vdots & \vdots & \ddots & \vdots & \vdots & \vdots \\
0 & 0 & \cdots & -t_{N_x-1,N_x} & t_{N_x-1,N_x} & 0 \\
\noalign{\vskip6pt}
0 & 0 & \cdots & 0 & -t_{N_x,N_x+1} & t_{N_x,N_x+1}
\end{array}\right)\, ,
\end{equation}
for Neumann boundary conditions, and
\begin{equation}
\fl
\bi{P}=\left(
\begin{array}{c}
p_1\\
\noalign{\vskip6pt}
p_2\\
\vdots\\
p_{N_x-1}\\
\noalign{\vskip6pt}
p_{N_x}
\end{array}\right)\, ,\qquad
\bi{B}_D=\left(
\begin{array}{c}
t_{0,1}P_i\\
\noalign{\vskip6pt}
0\\
\vdots\\
0\\
\noalign{\vskip6pt}
t_{N_x,N_x+1}P_f
\end{array}\right)\, ,\qquad
\bi{B}_N=\left(
\begin{array}{c}
t_{0,1}P_i+q_0\\
\noalign{\vskip6pt}
q_0\\
\vdots\\
q_0\\
\noalign{\vskip6pt}
q_0
\end{array}\right)\, ,
\end{equation}
in which $\bi{B}_D$ and $\bi{B}_N$ are the inhomogeneous terms for Dirichlet and Neumann boundary conditions, respectively. The calculations were carried out with the sparse matrix solver UMFPACK \cite{UMFPACK}, which can solve a sparse matrix equation in $O(N_x\log N_x)$ flops.


\section{Path Integral Formulation of Solutions to Darcy's Law}
\label{sec4}

The path integral provides an alternative approach to obtaining pressure statistics for flow in a random permeable medium. Introduced as an alternative formulation of quantum mechanics by Feynman in a seminal paper \cite{feynman}, the path integral is a weighted integral over all possible ``paths'' (in our case, pressure trajectories) for a noisy system (in our case, random permeabilities).   The path integral can be formulated for a broad range of physical problems that have a source of uncertainty or in noisy environments, most notably for quantum field theory \cite{huang}, statistical mechanics \cite{brush} and stochastic dynamics \cite{wio} and, as first pointed out by Wiener \cite{wiener}, to stochastic differential equations.  Here, we derive the path integral for Darcy's law by using using methods of classical statistical dynamics \cite{dedom,jouvet}. The methods described in Sec.~\ref{sec5} enable us to numerically evaluate the path integral to obtain pressure statistics and correlations once the statistics of the permeability have been specified. 

Our initial aim is to calculate the probability associated with some pressure path. Suppose we take a large number $N_x$ of pressure measurements $p(x_1),\ldots,p({N_x})$ at points separated by a small spatial interval $\delta x$ and subject to the boundary conditions $p(0)=P_i;~p({N_{x+1}})=P_f$, as illustrated in Fig.~\ref{fig4}(a). We have chosen to use DBC for this example because they are easy to visualize.  We denote the value that results from a pressure measurement at position $x_j$ by $p_j$.  A ``pressure path'' $\{p_j\}$ is defined by the values of the pressure across the system:
\begin{equation}
\{p_j\}=\{P_i,p_1,p_2,\ldots,P_f\}\, .
\label{eq14}
\end{equation}
The generating functional for expectations and correlation functions of the pressure that are determined by Darcy's law is, for a fixed log-permeability, expressed as an integral over pressure paths:
\begin{equation}
Z_{L}(\{u_i\})=\int\prod_i dp_i  \exp\biggl(\sum_iu_ip_i\biggr)\delta\biggl({p_i-p_{i-1}\over\delta x}+q_0 e^{-L_i}\biggr)\, .
\label{eq5}
\end{equation}
We have omitted the Jacobian $J=(\delta x)^{-N}$ from the argument of the $\delta$-function. A detailed calculation can be found in the article by Jouvet and Phythian \cite{jouvet}.  Although $J$ becomes infinite as $\delta x\to0$, this quantity is cancelled by the same divergence in the denominator in expressions for averages. Taking the average of $Z_L$ over the probability density of the log-permeability, we obtain the generating function for pressure correlations:
\begin{equation}
Z(\{u_i\})=\int \prod_i dL_i P(\{L_i\})Z_{L}(\{u_i\}) e^{-\sum_i L_i }\, .
\label{eq6}
\end{equation}
The factor $q_0$ in the Jacobian $q_0 \exp(-\sum_i L_i)$ has been omitted. The log-permeabilities are taken to follow a multivariate Gaussian distribution:
\begin{equation}
P(\{L_i\})={1\over (2\pi)^{N/2}|{\sf C}_L|^{1/2}}\exp\biggl[-\sum_{ij}L_i({\sf C}_L^{-1})_{ij}L_j\biggr]\, ,
\label{eq7}
\end{equation}
where ${\sf C}_L$ is the correlation matrix and $|{\sf C}_L|$ its determinant.  By substituting (\ref{eq5}) and (\ref{eq7}) into (\ref{eq6}) and again omitting constant prefactors, we obtain
\begin{eqnarray}
Z(\{u_i\})&=\int\prod_i dp_i\int\prod_i d{L_i}\exp\biggl(\sum_iu_ip_i\biggr)\exp\biggl[-\sum_{ij}L_i({\sf C}_L^{-1})_{ij}L_j \biggr]\nonumber\\
&\qquad\times\exp\biggl(-\sum_i L_i\biggr)\delta\biggl({p_i-p_{i-1}\over\delta x}+q_0 e^{-L_i}\biggr)\, .
\end{eqnarray}
By integrating over the $L_i$, we are left with a path integral for the probability density $Q$ of the pressures:
\begin{equation}
Q(\{p_i\},\{u_i\})={e^{-S(\{p_i\},\{u_i\})}\over Z}\, ,
\label{eq9}
\end{equation}
where
\begin{equation}
Z(\{u_i\})=\int\prod_i dp_i e^{-S(\{p_i\},\{u_i\})}\, ,
\end{equation}
with the discrete ``action''
\begin{eqnarray}
S(\{p_i\})&=\sum_{ij}\log\biggl({p_{i-1}-p_i\over q_0\delta x}\biggr)({\sf C}_L^{-1})_{ij}\log\biggl({p_{j-1}-p_j\over q_0\delta x}\biggr)\nonumber\\
&\quad+\sum_i\log\biggl({p_{i-1}-p_i\over q_0\delta x}\biggr)+\sum_i u_ip_i\, .
\label{eq10}
\end{eqnarray}
Averages over pressure are determined by logarithmic derivatives of $Z(\{u_i\})$:
\begin{equation}
\langle p_{k_1},p_{k_2},\ldots,p_{k_n}\rangle={\partial^n\log\bigl[Z(\{u_i\})\bigr]\over\partial u_{k_1}\partial u_{k_2}\cdots u_{k_n}}\bigg|_{\{u_i=0\}}\, .
\end{equation}
For example, the average $\langle p_k\rangle$ of the pressure $p_k$ at the $k$th lattice point is
\begin{equation}
\langle p_k\rangle={1\over Z}\int\prod_i dp_i\, p_k\, e^{-S(\{p_i\})}\, ,
\label{eq11}
\end{equation}
which confirms the cancellation of the omitted factors.   Higher-order correlation functions and cumulants are calculated analogously.

\section{Computational Methods for Evaluating Path Integrals}
\label{sec5}

The calculations are carried out on a spatial lattice with $N_x\gg 1$ elements (Fig.~\ref{fig4}). The computational task is to generate $N\gg 1$ pressure trajectories from the probability $e^{-S(\{p_i\})}$.  Once these trajectories are available, the calculation of the pressure statistics and correlations is straightforward. We use a Markov Chain Monte Carlo (MCMC) method, so called because the ``paths''  (\ref{eq14}) are updated to form a chain, within which each profile $\{p_j\}^{(\nu)}$ depends only on its predecessor $\{p_j\}^{(\nu-1)}$. The discrete ``action'' is given in (\ref{eq10}). Starting from an initial configuration $\{p_j\}^{(0)}$, typically an array of random numbers that respects the boundary conditions, we use the Metropolis-Hastings \cite{metropolis, hastings} (MH) algorithm to update the elements of this array.  The $N_x$ elements are normally updated in random order.  An element may be updated more than once, but will on average be visited once per ``sweep'' (series of $N_x$ updates). The MH algorithm is designed to generate paths from any probability distribution, using a symmetric proposal update. As applied to a specific lattice element $p_k$, the basic MH algorithm consists of four steps:

\begin{enumerate}

\item Generate a random number $u$ from a uniform probability distribution in the interval $[-h,h]$. The parameter $h$, known as the ``hit size''.

\item Propose a new value: $p_k^\prime=p_k+u$. If $h$ is too large, few changes will be accepted; too small and the exploration of the phase space will be slow.  An optimal acceptance rate (a conventional choice is 0.5) is obtained by adjusting $h$ after $N_x$ Metropolis updates according to the current acceptance rate. If the current acceptance rate is less than optimal, the hit size is decreased by one percent.  If the current acceptance rate is greater than optimal, the hit size is increased by one percent. 

\item Calculate the change $\Delta S$ in the action as a result of the proposed change. 

\item Accept the change with probability $\mathrm{Min}(1, e^{-\Delta S})$: changes that lower the action are always accepted.  A value of $p_k^\prime$ that would increase the action is accepted with probability $e^{-\Delta S}$.  

\end{enumerate}

\noindent
After one sweep the pressure profile is said to have been updated once. The MH algorithm satisfies the detailed-balance condition, which ensures reversibility of the chain in equilibrium \cite{morningstar}.
                                                                                                                                                                                                                                                                                                                                                                                                                           
Two further remarks are in order here. First, some initial number of measurements $N_{\mathrm{therm}}\gg 1$ must be discarded because $N_{\mathrm{therm}}$ sweeps are required to generate a path that is representative of the desired probability distribution.  Such a path is said to be ``thermalized''. Second, between every two paths used for measurements, some $N_{\mathrm{sep}}\gg 1$ must be abandoned because each path is created from a previous path, there is strong autocorrelation within the Markov chain.  Any set of paths is, therefore, representative of the probability law $e^{-(S\{p_i\})}$ only if a sufficient number of intermediate paths is discarded. 

With regard to the boundary conditions, $p_0=P_i$ and $q_0$ are fixed under NBC. The index 0 cannot chosen as part of the sweep. Under DBC, $p_0=P_i$ and $p_{N_x+1}=P_f$ are fixed.  The pressure is initially updated with only $p_0$ fixed; after $N_{\mathrm{sep}}$ updates $q_0$ is calculated and the pressure is rescaled accordingly.  Details of the MH algorithm and the calculation of path integrals on a lattice can be found in Ref. \cite{westbroek18a}.

In contrast to the finite-volume method, the path integral requires $O(N_x^2)$ flops to calculate a pressure realization. One factor $N_x$ arises from the number of lattice sites.  The number of required intermediate updates $N_{\mathrm{sep}}$  introduces a further factor $N_x$.  However,  the implementation of techniques such as overrelaxation \cite{creutz2, brown} and the multigrid method \cite{goodman} will likely decrease the run time considerably.
In addition, the performance of the Metropolis-Hastings algorithm can often be improved through directed sampling \cite{dir1,dir2,dir3}.

\section{Pressure Statistics for Neumann Boundary Conditions}
\label{sec6}

\begin{figure}
\centering
\hskip2cm\includegraphics[width=13cm]{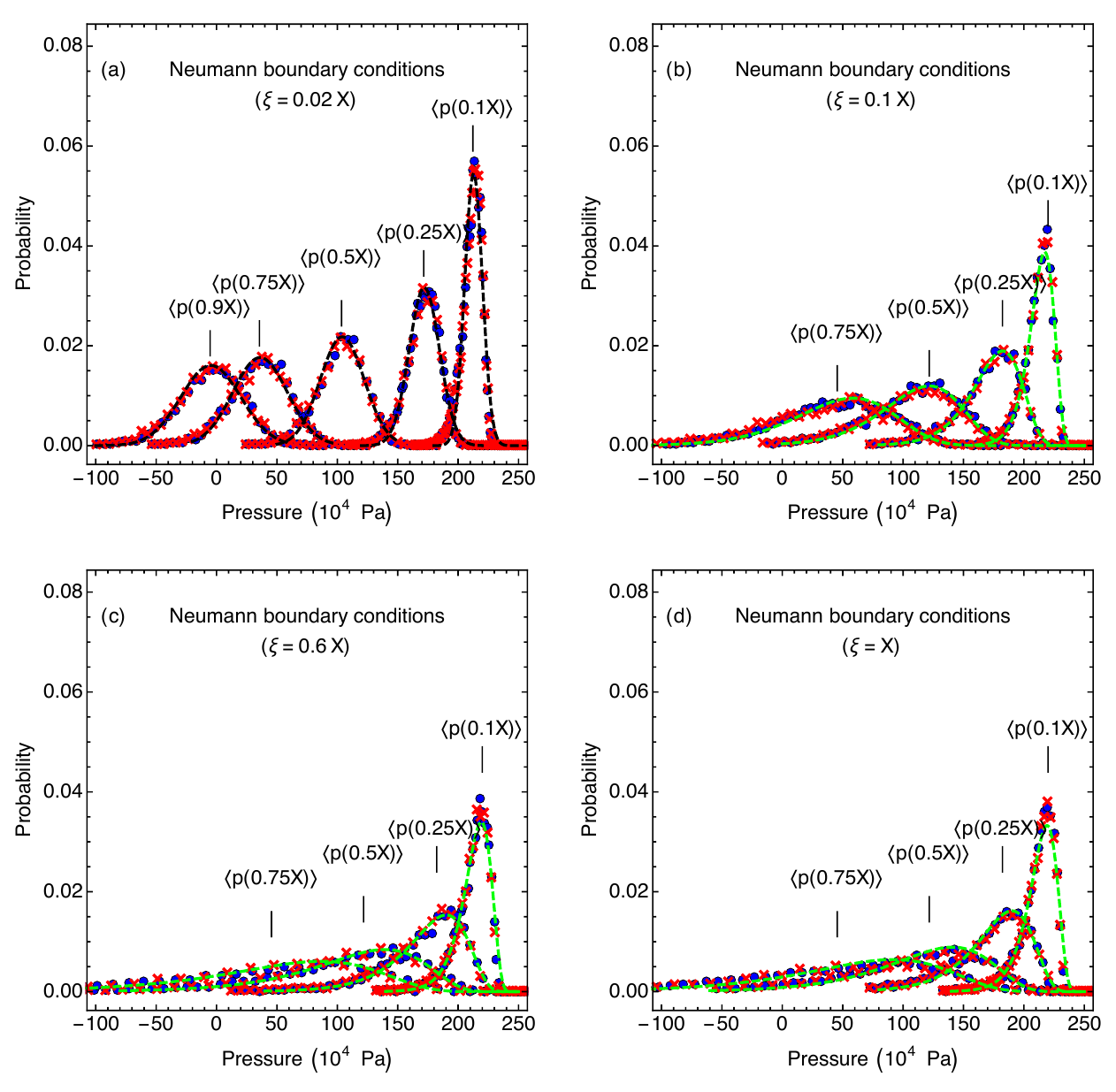}
\caption{Pressure statistics for Darcy flow under NBC for correlation lengths of the stochastic permeability (Fig.~\ref{fig2}) of (a) $\xi=0.02X$, (b) $\xi=0.1X$, (c) $\xi=0.6X$, and (d) $\xi=X$ obtained from the finite volume method (disks) and the path integral (crosses). Fewer data points are shown than used for statistical analysis. The dashed green and black lines represent the log-normal fits and Gaussian approximations, respectively. The calculations are based on $N=10^4$ simulations on lattices with $N_x=240$ sites and spacing $\Delta x=0.5~\mathrm{m}$. The path integral and stochastic data agree to a confidence level of 95\%.}
\label{fig5}
\end{figure}

Pressure statistics for Darcy flow obtained from the stochastic differential equation and the path integral are shown in Fig.~\ref{fig5}. We have used a system size of $X=240$~m with $P_i/X=10^4~\mathrm{Pa/m}$ and $q_0=10^{-6}~\mathrm{m/s}$, which are typical estimates for Darcy flow of oil in rocks.  The probability density of the pressure was calculated at various points in the rock for several values of the correlation length.  These correspond to vertical slices at the chosen points through the realizations of the type shown in Fig.~\ref{fig3}. Because the initial pressure is greater than the final pressure, the pressure distributions are shown at positions that increase from right to left. For the three largest values of the correlation length, the graph corresponding to $x=0.9X$ was left out to improve clarity. The pressure is fixed at $x=0$.  Away from this point, the distributions broaden in a manner determined by the correlation length of the permeability (Fig.~\ref{fig2}), but the free boundary condition at $x=X$ means that the broadening continues unabated for $x>0$.

Two types of distributions are used to fit the data in Fig.~\ref{fig2}:~log-normal and Gaussian. The equation 
\begin{equation}
\frac{dp(x)}{dx}=-\frac{q_0}{K(x)}
\end{equation}
can be interpreted as a steady-state Langevin equation for diffusion with log-normal noise. The resulting expression for the variance of the pressure,
\begin{equation}
\langle p^2(x)\rangle-\langle p(x)\rangle^2
= q_0^2 e^{\sigma^2}\int_0^x \int_0^x \Bigl[e^{\sigma^2 e^{-|u-v|/\xi}}-1\Bigr]\,du\,dv\, ,
\end{equation}
demonstrates the vanishing influence of the correlation length with increasing $x/\xi$.
For NBC, Darcy's law takes the form (\ref{DarcyNBC}), where $R(x)$, given in (\ref{eq:R}), is the source of noise.


The resistance term $R(x)$ is an integral over correlated log-normal stochastic variables. For small values of $x/\xi$, $R$ is the integral over weakly correlated log-normal variables over a short distance (small $\xi$), or the integral over more strongly correlated stochastic variables over a longer distance (larger $\xi$).  In both of these cases, we are able to make a log-normal fit to $R(x)$ and, consequently, to $p(x)$.  These fits (Fig.~\ref{fig5}) are based on the empirical first and second moments of the simulations. 

\begin{table}
\caption{\label{table1} Means and standard deviations of the pressure as a function of position $x$ and noise correlation length $\xi$ for Neumann boundary conditions for a one-dimensional system obtained from simulations ($\mu_s$ and $\sigma_s$) and calculations from the solution to the Fokker--Planck equation ($\mu_c$ and $\sigma_c$), and from a calculation that incorporates the discreteness of the simulation ($\sigma_d$). The units of $\xi$ and $x$ are $X=240$~m.  Means and standard deviations are expressed in units of $10^4$~Pa.}
\begin{indented}
\lineup
\item[]\begin{tabular}{@{\hskip6pt}lllllll}
\br
$\xi (X)$ & $x (X)$ & \multicolumn{1}{c}{$\mu_s$} & \multicolumn{1}{c}{$\mu_c$} & \multicolumn{1}{c}{$\sigma_s$} & \multicolumn{1}{c}{$\sigma_d$} & \multicolumn{1}{c}{$\sigma_c$}\\
\mr
&0.1&212.168&212.804&\0\08.08306&\0\07.32955&\0\07.97994\\
&0.25&171.354&172.011&\013.5634&\012.6210&\013.4854\\
0.02&0.5&103.348&104.022&\019.5774&\018.3138&\019.4676\\
&0.75&\035.4959&\036.0333&\024.3331&\022.6163&\024.0024\\
&0.9&\0\0\-5.27945&\0\0\-4.76007&\026.6637&\024.8427&\026.3513\\ \ms
&0.1&212.318&212.804&\012.5381&\012.1807&\012.2510\\
&0.25&171.466&172.011&\025.5700&\025.1552&\025.1849\\
0.1&0.5&103.167&104.022&\039.9844&\039.8835&\039.8997\\
&0.75&\035.3963&\036.0333&\050.5274&\050.7368&\050.7478\\
&0.9&\0\0\-5.26133&\0\0\-4.76007&\055.8586&\056.2633&\056.2723\\ \ms
&0.1&212.055&212.804&\014.6806&\013.9840&\014.0545\\
&0.25&171.198&172.011&\034.4524&\033.5988&\033.6267\\
0.6&0.5&103.320&104.022&\063.2270&\062.8236&\062.8375\\
&0.75&\035.2116&\036.033&\088.2456&\088.5658&\088.5751\\
&0.9&\0\0\-5.83536&\0\0\-4.76007&102.159&102.640&\0102.648\\ \ms
&0.1&212.203&212.804&\014.6486&\013.8463&\014.2264\\
&0.25&171.397&172.011&\035.1531&\034.2714&\034.6159\\
1.0&0.5&103.082&104.022&\067.2820&\065.9685&\066.3126\\
&0.75&\034.7882&\036.033&\096.7118&\095.1455&\095.4987\\
&0.9&\0\0\-6.18528&\0\0\-4.76007&113.349&111.588&111.947\\
\br
\end{tabular}
\end{indented}
\end{table}

Table~\ref{table1} compares the mean and standard deviation of the pressure distribution obtained from the evaluation of the path integral ($\mu_s$ and $\sigma_s$), from the solutions (\ref{Rexpcont}) and (\ref{eqC5}) of the Fokker--Planck equation of the Ornstein-Uhlenbeck process ($\mu_c$ and $\sigma_c$), and the corresponding discrete forms (\ref{Rexpdisc}) and (\ref{Rvardisc}) corresponding to the discretization of the finite-volume method in Sec.~\ref{sec3} ($\mu_d=\mu_c$ and $\sigma_d$).  There are two sources of the discrepancies between the results from the path integral, which are regarded as ``exact'', and the two sets of calculations in \ref{secA3}:~lattice effects and correlations.  Figure~\ref{fig5} shows that the Gaussian approximation is most accurate for the smallest correlation length (Fig.~\ref{fig5}(a)), with correlation effects becoming more evident with increasing correlation length (Fig.~\ref{fig5}(b,c,d)).  Correlation effects play a role in both the discrete and the continuous approximations to the path integral calculations. The Gaussian approximation becomes more appropriate as the ratio $\xi/x$ decreases. The ``breakdown of Gaussianity'' can be clearly seen in Fig.~5(b,c,d):~Gaussian fits are only appropriate for the smallest correlation length. Due to the strong correlation, the pressure distributions are seen to broaden appreciably away from the boundary. The data in Table~\ref{table1} support this trend in the comparisons between the values of the mean and standard deviation obtained from the path integral and the approximations. 

A comparison between the discrete and continuous approximations to the standard deviation ($\sigma_d$ and $\sigma_c$, respectively) makes apparent that $\sigma_c$ is a better approximation to the standard deviation $\sigma_s$ obtained from the path integral calculation.
These discrepancies are indicative of the error caused by the discretization, which diminish with increasingly refined lattice spacing.

We now consider the case of small $\xi/x$. In the limiting case $\xi \to 0$, each stochastic variable $K(x)$ is drawn from the Gaussian probability function (\ref{secA1}). Indeed, the central limit theorem (CLT) mandates that the sum of $N$ independent, identically distributed random variables tends to a normal distribution in the limit of large $N$ if all moments of the distribution are finite. An alternative version of the CLT holds for correlated random variables and states that the sum of $N$ realizations of an ergodic process (whose long-term average is equal to its expectation value \cite{vK}), will behave as a Gaussian random variable in the limit of large $N$ \cite{bouchaud}. The process $K$ is a Markov process with continuous trajectories, otherwise known as a ``diffusion'' process \cite{pavliotis}.  The probability law of $K$ is invariant under time reversal. 
Since all reversible diffusions are ergodic \cite{pavliotis}, the alternative version of the CLT applies to $R(x)$. 

Under what conditions can $R(x)$ be approximated by a Gaussian random variable? Clearly, the criterion depends on the ratio $\xi/x$.  We have determined the order of magnitude of $\xi/x$ below which $R(x)$ is approximately Gaussian.  To do so, we made use of the Kolmogorov--Smirnov (KS) test, which is a statistical test that determines the probability (or ``$p$-value'' \cite{wasserstein}) that a data sample follows a given distribution (one-sided test), as well as the probability that two data sets follow the same distribution (two-sided test) \cite{KSpaper}. For $R(x)$, we performed a one-sided KS test at the $95\%$ confidence level:~if the $p$-value is below $5\%$, we reject the ``null hypothesis'' that the sample is normally distributed. Of course, the greater the $p$-value, the better the quality of the approximation.  A two-sided KS test was carried out for all pairs of data sets (FVM and path integral) at the $95\%$ confidence level, which all pairs of data sets passed.

\begin{figure}
\centering
\includegraphics[width=0.4\textwidth]{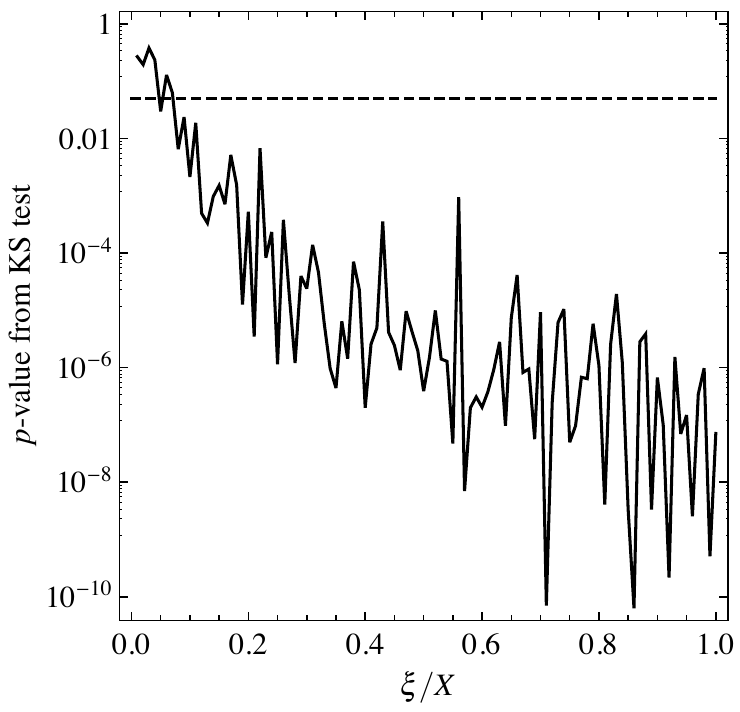}
\caption{The $p$-value obtained from the KS test for $R(X)$ as a function of $\xi/X$.}
\label{fig6}
\end{figure}

We have carried out a KS test for $100$ different values of $\xi/x$, each based on $N=1000$ realizations of $R$ \cite{KS}. The realizations were of $R(X)$;~if $R$ is found to be Gaussian at any point in $[0,X]$, then it is Gaussian on the entire interval, due to the strict stationarity of Gaussian stochastic processes (\ref{secA1}).  From Fig.~\ref{fig6} we infer that the Gaussian approximation breaks down for 
\begin{equation}
\label{Gausscondition}
x\gtrsim 10\,\xi.
\end{equation}
For $\xi=0.02 X$, we were therefore able to make Gaussian fits to $p(x)$ for all values of $x$. These fits are shown in Fig. \ref{fig6}.  If it is possible to make a Gaussian approximation, it is advantageous to do so, because it can be based on the theoretical mean and variance of $p(x)$ and does not require any simulations. A calculation of the first and second moments of $p(x)$ under NBC can be found in \ref{NBCGauss}.


\section{Pressure Statistics for Dirichlet Boundary Conditions}
\label{sec7}

Under Dirichlet boundary conditions, the pressure $p(x)$ takes the form
\begin{equation}
p(x)=P_i-(P_f-P_i)\frac{R(x)}{R(X)}.
\end{equation}
From the Kolmogorov--Smirnov test in Sec.~\ref{sec6}, we know that $R(x)$ can be approximated by a Gaussian random variable if the condition (\ref{Gausscondition}) is met. 
The ratio 
\begin{equation}
\label{Rfrac}
\frac{R(x)}{R(X)}=\frac{\int_0^x \frac{1}{K(x')}dx'}{\int_0^x \frac{1}{K(x')}dx'+\int_x^X \frac{1}{K(x')}dx'}
\end{equation}
is then a function of correlated Gaussians.  We have taken as our working assumption that a Gaussian approximation can be made to (\ref{Rfrac}) if $x\gtrsim 10\xi$. This assumption, verified by further Kolmogorov--Smirnov tests, turns out to be correct. As for NBC, we have made Gaussian approximations to the pressure distributions for the parameter choice $\xi=0.02X$;~for the other parameter choices, we have made log-normal fits to the pressure distributions. A derivation of these approximations is given in \ref{DBCGauss}.


\begin{figure}
\centering
\hskip2cm\includegraphics[width=13cm]{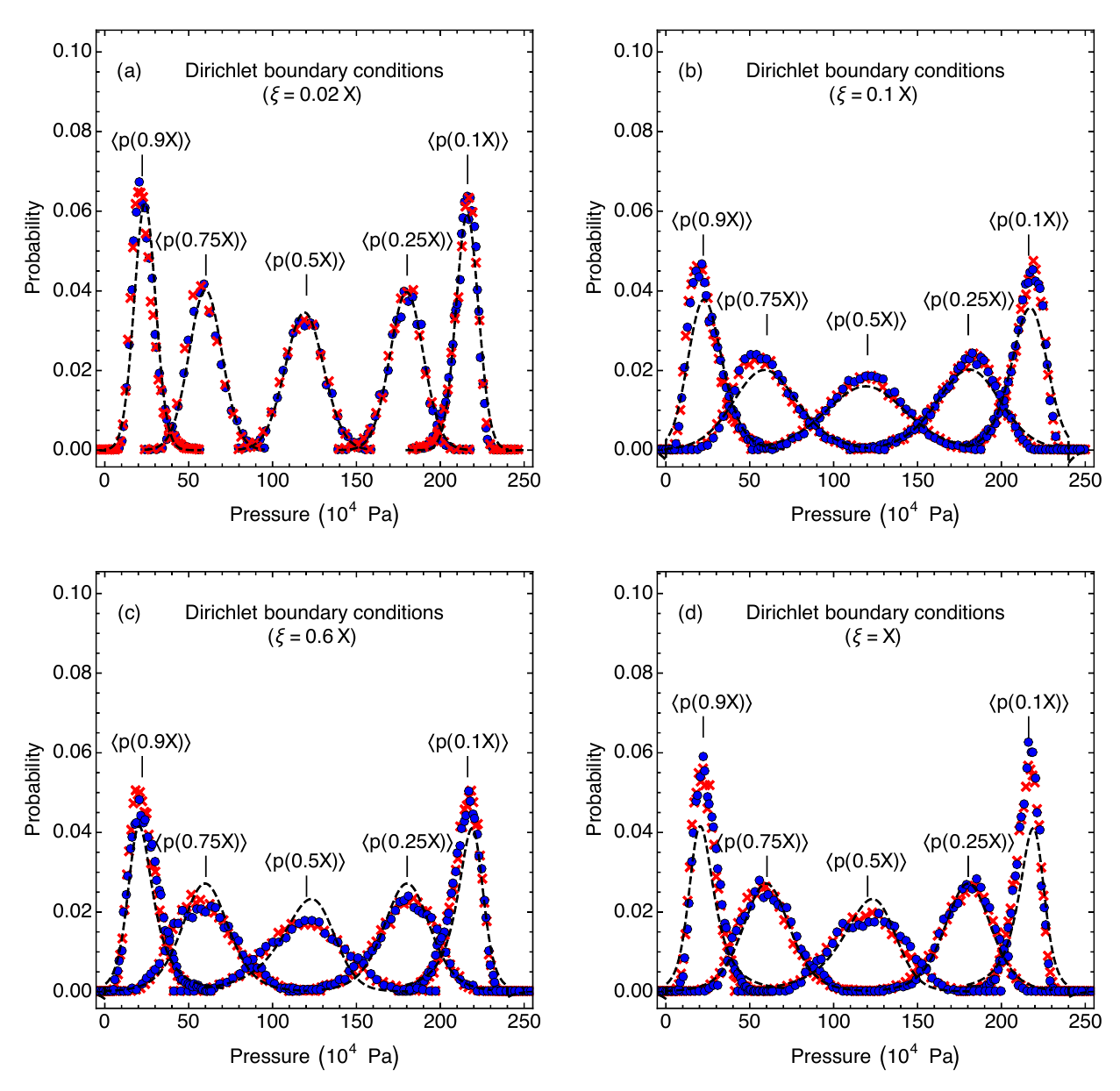}
\caption{Pressure statistics for Darcy flow under DBC for correlation lengths $\xi=0.02X$ (a), $\xi=0.1X$ (b), $\xi=0.6X$ (c) and $\xi=X$ (d). Simulations were done using the finite volume method (disks) and path integral method (crosses). Fewer data points are shown than used for statistical analysis. The dashed green and black lines represent the log-normal fits and Gaussian approximations, respectively.
The calculations are based on $N=10^4$ simulations on a lattice of $N_x=240$ lattice sites and lattice spacing $\Delta x=0.5~\mathrm{m}$. 
The path integral and stochastic data agree to a confidence level of 95\%.}
\label{fig7}
\end{figure}


The Gaussian theoretical curves and log-normal fits are shown in Fig.~\ref{fig7}. Other than the final pressure, $P_f/X=0$~Pa/m, and $q_0$, which is not fixed, all parameters were set to the same values as under Neumann boundary conditions. Because the initial and final pressures are fixed, the distributions are narrowest near the ends of the system. The distributions broaden away from the endpoints. The pressure range is greater for the system with the smaller correlation length for the permeability. This results from the pressure paths showing a smaller variation with the smaller correlation length.  In the case of DBC, the pressure distributions $p(x)$ are symmetric about $x=X/2$, due to the strict stationarity of the Ornstein-Uhlenbeck process (\ref{secA1}).  The influence of the correlation length is, therefore, felt at both ends of the interval $[0,X]$, and the Gaussian approximation is best in the center of the ``rock''. 

\begin{table}[t]
\caption{\label{table2}Means and standard deviations of the pressure as a function of position $x$ and noise correlation length $\xi$ for Dirichlet boundary conditions for a one-dimensional system obtained from simulations ($\mu_s$ and $\sigma_s$) and calculations from the solution to the Fokker--Planck equation ($\mu_c$ and $\sigma_c$), and from a calculation that incorporates the discreteness of the simulation ($\mu_d$ and $\sigma_d$). The units of $\xi$ and $x$ are $X=240$~m.  Means and standard deviations are expressed in units of $10^4$~Pa.}
\begin{indented}
\lineup
\item[]\begin{tabular}{@{\hskip6pt}llllllll}
\br
\multicolumn{1}{c}{$\xi\,(X)$}&\multicolumn{1}{c}{$x\,(X)$}&\multicolumn{1}{c}{$\mu_s$}&\multicolumn{1}{c}{$\mu_d$}&\multicolumn{1}{c}{$\mu_c$}&\multicolumn{1}{c}{$\sigma_c$}&\multicolumn{1}{c}{$\sigma_d$}&\multicolumn{1}{c}{$\sigma_c$}\\
\mr
&0.1&215.537&215.638&215.793&\06.65118&\07.39637&\08.05491\\
&0.25&179.497&179.830&179.748&\09.976593&\08.61085&10.4318\\
0.02&0.5&119.300&120.50&119.499&11.76703&11.2895&12.0930\\
&0.75&\059.3505&\059.2513&\059.2475&10.1345&\09.64244&10.3687\\
&0.9&\023.2585&\023.1188&\023.1126&\06.512375&\06.08723&\07.71568\\ \ms
&0.1&215.517&212.630&214.834&\09.782208&10.7666&\09.08430\\
&0.25&179.703&178.413&179.655&17.73364&19.0602&14.6467\\
0.1&0.5&119.857&120.427&119.500&21.855&26.1954&17.1548\\
&0.75&\060.0631&\059.8698&\059.3168&17.65783&20.9629&14.4449\\
&0.9&\023.8099&\023.7858&\023.2374&\09.600161&11.0541&\08.66795\\ \ms
&0.1&214.934&204.426&203.015&\08.798261&22.7199&25.6091\\
&0.25&178.866&169.881&168.652&17.41341&35.2504&36.1574\\
0.6&0.5&119.463&116.881&115.825&22.56851&44.9074&45.2879\\
&0.75&\060.1718&\062.4917&\062.4303&17.27851&35.2585&35.6411\\
&0.9&\024.0128&\026.8928&\026.9697&\08.472904&21.3853&21.6713\\ \ms
&0.1&215.072&202.156&200.700&\07.567443&26.8830&29.5648\\
&0.25&178.946&166.939&166.407&15.25785&38.9908&39.8613\\
1.0&0.5&119.497&115.702&114.638&19.99505&47.5015&47.8495\\
&0.75&\059.8951&\062.6097&\062.5277&15.08735&37.3308&37.6855\\
&0.9&\023.817&\027.4292&\027.4930&7.327701&23.4083&23.6594\\
\br
\end{tabular}
\end{indented}
\end{table}

The comparison between the mean and standard deviation obtained from path integral simulations with those from discrete and continuous calculations with the Gaussian approximation is shown in Table~\ref{table2}.   As expected, the accuracy of the Gaussian approximation diminishes with increasing correlation length. The spatial effect of increased Gaussianity is less apparent under Dirichlet boundary conditions:~the ``rock'' has two boundaries.  One can only get half as far away from the boundary, where the approximation is most fitting.

Contrary to the calculations done for Neumann boundary conditions, there is no exact equality between the mean resulting from the Fokker-Planck equation ($\mu_s$) and its discrete counterpart ($\mu_d$).  Under Neumann boundary conditions, the approximate mean depends only on the expectation value of $R$ (Eq.~(\ref{c2})). This quantity does not depend on the nature of the approximation. By contrast, under Dirichlet boundary conditions the standard deviation of the pressure depends on many quantities, including the variance of $R$ and the correlation between values of $R$ at different points.  These variables in turn depend on correlation effects, which a discrete setup estimates more crudely than a continuous one.

\section{Conclusions and Future Work}
\label{sec8}

This study has shown that a path integral formulation of Darcy's law can be used to calculate pressure statistics for flow through a one-dimensional permeable medium.  The results of these calculations agree with those acquired through conventional finite-volume techniques.  A Gaussian approximation, whose advantage lies in the small computational effort, is applicable when the correlation length is much smaller than the system size.  For larger correlation lengths, the pressure distributions are best approximated by a log-normal distribution. 

The one-dimensional calculations presented here are intended to demonstrate the accuracy and viability of the path integral method and to gain insight into the pressure statistics of constrained Darcy flow through a porous medium. In fact, the one-dimensional system has relevance to actual physical settings because Darcy flow is often constrained to flow along one main direction, with negligible flow in the lateral directions.  Darcy flow in two and three dimensions requires an extension of the method described here.  A full description of this extension and the results of two-dimensional simulations will be published in a forthcoming article. 

Apart from the extension of our method to higher dimensions, there are two other main areas where the path integral affords a more general description of Darcy flow.  The lognormal distribution of the permeabilities (\ref{eq7}) can be replaced by any distribution \cite{dedom,phythian77};~the only effect of such a replacement on our implementation would be to modify the acceptance/rejection criteria in the Metropolis--Hastings method because of the modified Lagrangian.

The other extension, to multiphase flow, is a more substantial endeavor.  Multiphase flow through porous media is important for a various applications, such as CO$_2$ sequestration,\cite{juanes12} and enhanced oil recovery \cite{orr84}. These often involve the displacement of a nonwetting invading fluid from a porous medium by a wetting fluid (imbibition). There are several formulations of the equations of multiphase flow, with the most realistic expressed in terms of coupled stochastic nonlinear partial differential equations.  Casting the solutions to these equations as a path integral and adapting the MCMC method for the evaluation of correlation functions will be the subject of future work.

\ack
MJEW was supported through a Janet Watson scholarship from the Department of Earth Science and Engineering and a studentship in the Centre for Doctoral Training on Theory and Simulation of Materials funded by the EPSRC (EP/L015579/1), both at Imperial College London. 
GAC was supported by a studentship in the CDT in TSM, funded by the EPSRC (EP/G036888/1), for the duration of his time at Imperial College London.

\newpage

\appendix

\section{Conditional Probability for Log-Permeability}
\label{secA1}

The conditional probability density $P(L_2,x_2;L_1,x_1)$ that, given that the log permeability takes the value $L_1$ at $x_1$, the value $L_2$ at $x_2$ is
\begin{equation}
\fl P(L_2,x_2;L_1,x_1)={1\over\sqrt{2\pi\sigma^2[1-e^{-(x_2-x_1)/\xi}}]}\exp\left\{-{1\over2\sigma^2}{[L_2-L_1e^{-(x_2-x_1)/\xi}]^2\over 1-e^{-2(x_2-x_1)/\xi}}\right\}\, .
\label{eqA1}
\end{equation}
This function is a Gaussian probability density with mean $L_1e^{-(x_2-x_2)/\xi}$ and variance $\sigma^2[1-e^{-2(x_2-x_1)/\xi}]$, where $\xi$ is a correlation length. The initial condition for $P$, when $x_2=x_1$, is
\begin{equation}
P(L_2,x_1;L_1,x_1)=\delta(L_2-L_1)\, ,
\end{equation}
and, when $(x_2-x_1)/\xi\to\infty$, that is, when $x_2-x_1\gg\xi$, $P$ approaches a Gaussian distribution for $L_2$:
\begin{equation}
\lim_{(x_2-x_1)/\xi\to\infty}\hskip-0.7cmP(L_2,x_2;L_1,x_1)={1\over\sqrt{2\pi\sigma^2}}\exp\left(-{L_2\over2\sigma^2}\right)\, .
\end{equation}
In fact, $P$ in (\ref{eqA1}) is the Green's function for
\begin{equation}
{\partial P\over\partial x}={1\over\xi}{\partial(\ell P)\over\partial\ell}+{\sigma^2\over\xi}{\partial^2P\over\partial\ell^2}\, ,
\end{equation}
which is the Fokker--Planck equation for an Ornstein--Uhlenbeck process with drift $1/\xi$ and diffusion $\sigma^2/\xi$.

\section{Correlation Functions for Permeability}
\label{secA2}

With the initial value $L_1$ in (\ref{eqA1}) drawn from  a Gaussian probability density with mean zero and variance $\sigma^2$,
\begin{equation}
P_0(L_1)={1\over\sqrt{2\pi\sigma^2}}\exp\left(-{L_1^2\over2\sigma^2}\right)\, ,
\end{equation}
the expectation value  $\mathbb{E}(K(x))$, where $K(x)=e^{L(x)}$, is calculated as
\begin{equation}
\mathbb{E}(K(x))=\int_{-\infty}^\infty\int_{-\infty}^\infty e^L\, P(L,x;L_0,0)P_0(L_0)\, dL\, dL_0=e^{\sigma^2/2}\, .
\label{eqB2}
\end{equation}
The two-point correlation function $\mathbb{E}(K(x_1)K(x_0))$ is
\begin{eqnarray}
\mathbb{E}(K(x_2)K(x_1))&=&\int_{-\infty}^\infty\int_{-\infty}^\infty e^{L_2}e^{L_1}\, P(L_2,x_2;L_1,x_1)P_0(L_1)\, dL_1\, dL_2\nonumber\\
\noalign{\vskip3pt}
&=&\exp\left\{\sigma^2\left[1+e^{-(x_2-x_1)/\xi}\right]\right\}\, ,
\end{eqnarray}
from which we obtain
\begin{eqnarray}
\mbox{Cov}(K(x_1)K(x_0))&=&\mathbb{E}(K(x_1)K(x_0))-\mathbb{E}(K(x_1))\mathbb{E}(K(x_0))\nonumber\\
\noalign{\vskip3pt}
&=&e^{\sigma^2}\left\{\exp\left[1+e^{-(x_1-x_0)/\xi}\right]-1\right\}\, .
\label{eqB4}
\end{eqnarray}
The mean and covariance in (\ref{eqB2}) and (\ref{eqB4}) are clearly invariant under translations of $x$. Finally, since $K^{-1}(x)=e^{-L(x)}$, the probability distribution for $K^{-1}$ is identical to that of $K$.

\section{Gaussian Approximations}
\label{secA3}

\subsection{Neumann boundary conditions}\label{NBCGauss}

From Darcy's law and the definition of $R$ in (\ref{eq:R}), we obtain 
\begin{equation}
p(x)=P_i-q_0 R(x)\, ,
\end{equation} 
where the initial pressure $P_i=p(0)$.  For Neumann boundary conditions, we specify $P_i$ and $q_0$. 
If the correlation length is small, we can use the following Gaussian approximation for $p(x)$:
\begin{equation}\label{c2}
p(x)\sim \mathcal{N}(\mu,\sigma^2)=\mathcal{N} \left(P_i-q_0 \mathbb{E}[R(x)],~ q_0^2\mbox{Var}(R(x))\right).
\end{equation} 
The first and second moments of $R(x)$ are calculated based on those of the permeability in (\ref{eqB2}) and (\ref{eqB4}): 
\begin{eqnarray}
\mathbb{E}[R(x)]&=&\int_0^x e^{\sigma^2/2}~dx=e^{\sigma^2/2}x \label{Rexpcont} \\
\mathrm{Var}[R(x)]&=&e^{\sigma^2}\left\{ \int_0^x \int_0^x e^{\sigma^2 e^{-|x'-x''|/\xi}}-1\,dx'\,dx''\right\}. \label{Rvarcont}
\end{eqnarray}
Similarly, the covariance of $R$ is given by:
\begin{equation}
\mathrm{Cov}[R(x),R(y)]=e^{\sigma^2}\left\{ \int_0^x \int_0^y e^{\sigma^2 e^{-|x'-x''|/\xi}}-1\,dx'\,dx''\right\}.
\label{eqC5}
\end{equation}

The expressions in (\ref{Rexpcont}), (\ref{Rvarcont}), and (\ref{eqC5}) represent statistical characteristics of the permeability associated with a continuous medium.  However, the evaluation of the path integral in Sec.~\ref{sec4} is carried out on a lattice with a particular lattice spacing.  Thus, for consistency, comparisons between the pressure statistics obtained from the discrete path integral and those calculated directly from the permeability distribution function should be based on discrete approximations to the first and second moments of $R$.  Referring to Fig.~\ref{fig4}, we arrive at the following discrete definition of $R$:
\begin{equation}
R(l)/\delta x=
\left\{\begin{array}{cll}
0 &\mbox{if}&l=0 \\
\noalign{\vskip6pt}
\frac{1}{2}K(1)+\sum_{i=1}^{l-1} K(i) &\mbox{if}&0<l<N_x \\
\noalign{\vskip6pt}
\frac{1}{2}K(1)+\sum_{i=1}^{l-1} K(i) + \frac{1}{2}K(N_x-1) &\mbox{if}& l=N_x.
\end{array}\right.
\end{equation}
We concentrate here on the most relevant case: $0<l<N_x$. 
The expectation value is then,
\begin{equation}\label{Rexpdisc}
\mathbb{E}[R(l)/\delta x]=\frac{1}{2}e^{\sigma^2/2} + (l-1) e^{\sigma^2/2} = \left(l-\frac{1}{2}\right) e^{\sigma^2/2},
\end{equation}
and the second moment of $R$ is
\begin{eqnarray}\label{Rsq}
\mathbb{E}[R(l)^2/\delta x^2]&=&\mathbb{E}\left[\frac{1}{4}K(1)^2+\sum_{i=1}^{l-1}[K(i)]+\sum_{i=1}^{l-1}\sum_{j=1}^{l-1}[ K(i) K(j)]\right] \nonumber \\
&=&\frac{1}{4} e^{\sigma^2} + \sum_{i=1}^{l-1}\left[e^{\sigma^2} e^{1+e^{-1/\xi}}\right] 
+(l-1)e^{\sigma^2}\nonumber\\
&& + 2\sum_{p=1}^{l-1}\left[ (l-1-p) e^{\sigma^2 \left(1+e^{1+e^{-p/\xi}}\right)}\right].
\end{eqnarray}
Subtracting the square of (\ref{Rexpdisc}) from (\ref{Rsq}), we obtain an expression for the discrete variance of $R$:
\begin{eqnarray}\label{Rvardisc}
\mbox{Var}[R(l)/\delta x]&=&\left(l-\frac{3}{4}\right) e^{\sigma^2} + \sum_{i=1}^{l-1}\left[e^{\sigma^2\left(1+e^{-i/\xi}\right)}\right] \nonumber \\
&&+2\sum_{p=1}^{l-1} \left[ (l-1-p) e^{\sigma^2\left(1+e^{-p/\xi}\right)}\right] - \left(l-\frac{1}{2}\right)^2 e^{\sigma^2}.
\end{eqnarray}

The Gaussian curves in Fig.~\ref{fig5} are based on the discrete definition of $R$.  The continuum definition, $R(x)=\int_0^x e^{-L(x')}~dx'$, gives rise to the same expectation values, but smaller variances. The discrepancies are due to the discretization of the lattice: in the limit $\delta x \to 0$, the expressions (\ref{Rvardisc}) and (\ref{Rvarcont}) give rise to the same numerical values. Note, in addition, that $R$ is strictly positive.  Because the Gaussian approximation is symmetric, the positivity of $R$ makes for a poor agreement between data and approximation for small values of $x$, especially for large values of the correlation length.

\subsection{Dirichlet boundary conditions}
\label{DBCGauss}

Again, we start from the integrated version of Darcy's law (\ref{pa}).
In the Dirichlet formulation, the initial pressure $P_i$ and final pressure $P_f=p(X)$ are fixed.  Then,
\begin{equation}
q_0=-\frac{P_f-P_i}{R(X)}.
\end{equation}
If we define $\Delta p\equiv P_i-P_f$, the pressure $p(x)$ is given by
\begin{equation}
p(x)=P_i-\Delta p \frac{R(x)}{R(X)}.
\end{equation}
To approximate $p(x)$ by a Gaussian random variable, we must find a Gaussian approximation to the ratio $R(x)/R(X)$, which is of the form
\begin{equation}
Z\equiv \frac{X}{X+Y},
\end{equation}
for stochastic variables 
\begin{equation}\label{XY}
X=\int_0^x \frac{1}{K(x')}~dx'\, ,\quad Y=\int_x^X \frac{1}{K(x')}~dx'\, .
\end{equation}
For this choice of stochastic variables in the regime $\xi \gtrsim 0.1X$, which defines the domain of applicability of the Gaussian approximation, we can assume that the mean is much greater than the variance.  This assumption is needed to ensure that the fraction $Z$ is strictly positive, and the cumulative distribution function of $Z$ can be defined sensibly. We denote the means of $X$ and $Y$ by $\mu_x$ and $\mu_y$ and the variances by $\sigma_x^2$ and $\sigma_y^2$, respectively, so the Gaussian forms of $X$ and $Y$ are
\begin{equation}\label{GaussianAssumption}
X \sim \mathcal{N}(\mu_x, \sigma_x^2)\, ,\qquad
Y \sim \mathcal{N}(\mu_y, \sigma_y^2).
\end{equation}
For Darcy's law, (\ref{GaussianAssumption}) translates to
\begin{equation}\label{RAssumption}
\begin{array}{rll}
R(x) &\hskip-6pt \sim &\hskip-6pt  \mathcal{N}(\mathbb{E}[R(x)], \mathrm{Var}[R(x)]) \\
\noalign{\vskip3pt}
R(X-x) &\hskip-6pt \sim &\hskip-6pt  \mathcal{N}(\mathbb{E}[R(X-x), \mathrm{Var}[R(X-x)]), 
\end{array}
\end{equation}
as given in Equations (\ref{Rexpcont}) and (\ref{Rvarcont}). 
Equation (\ref{RAssumption}) holds for $x\leq{1\over2} X$. If $x > {1\over2}X$, the condition
\begin{equation}
\frac{X}{X+Y}\leq z
\end{equation}
becomes
\begin{equation}
\frac{Y}{Y+X} \leq 1-z.
\end{equation}
The discussion below still follows, with the renamed variables. 
The covariance matrix of $X$ and $Y$ is symbolically denoted by
\[ \left( \begin{array}{cc}\label{covXY}
\sigma_x^2 & c_{xy}  \\
c_{xy} & \sigma_y^2 \end{array} \right),\]
where $c_{xy}\equiv \rho\,\sigma_x\sigma_y$ is the covariance matrix of $X$ and $Y$.
The symmetric, real, positive-definite covariance matrix admits the Cholesky decomposition
\[ \left( \begin{array}{cc}
\sigma_x & 0  \\
\sigma_y\, \rho & \sigma_y\sqrt{1-\rho^2} \end{array} \right).\]
The sum $X+Y$ can be written as a linear combination of unit normal ($\sim\mathcal{N}(0,1)$) random variables $U$ and $V$. 
We observe that
\begin{eqnarray}
\fl\left(\begin{array}{cc}
X & Y
\end{array}\right)
\left(\begin{array}{c}
X\\
\noalign{\vskip6pt}
Y
\end{array}\right)&=&
\left(\begin{array}{cc}
U & V
\end{array}\right)
\left(\begin{array}{cc}
\sigma_x^2 & \rho\,\sigma_x\sigma_y\\
\noalign{\vskip6pt}
\rho\,\sigma_x\sigma_y & \sigma_y^2
\end{array}\right)\left(\begin{array}{c}
U\\
\noalign{\vskip6pt}
V
\end{array}\right)\nonumber\\
\noalign{\vskip3pt}
&=&\left(\begin{array}{cc}
U & V
\end{array}\right)
\left(\begin{array}{cc}
\sigma_x & 0\\
\noalign{\vskip6pt}
\rho\,\sigma_y & \sigma_y\sqrt{1-\rho^2}
\end{array}\right)
\left(\begin{array}{cc}
\sigma_x & \rho\,\sigma_y\\
\noalign{\vskip6pt}
0 & \sigma_y\sqrt{1-\rho^2}
\end{array}\right)
\left(\begin{array}{c}
U\\
\noalign{\vskip6pt}
V
\end{array}\right)\nonumber\, .\\
\end{eqnarray}
Therefore, $X+Y$ has the same distribution as 
\begin{equation}
\mu_x+\mu_y+(\sigma_x+\rho \sigma_y)U+\sqrt{1-\rho^2}\sigma_y V.
\end{equation}

We can now find an expression for the cumulative distribution function of $\frac{X}{X+Y}\equiv Z$:
\begin{eqnarray}
F(z)&=&\mathbb{P}(Z\leq z) \nonumber\\
&=&\mathbb{P}\left\{ \frac{[(1-z)\mu_x-z\mu_y]\sigma_x+[(1-z)\sigma_x^2-z c_{xy}]U}{z\sqrt{\sigma_x^2 \sigma_y^2-c_{xy}}}<V\right\}\nonumber\\
&=&\mathbb{P}\left\{\frac{[(1-z)\mu_x -z\mu_y]\sigma_x}{z\sqrt{\sigma_x^2\sigma_y^2-c_{xy}}}+\frac{[(1-z)\sigma_x^2-z c_{xy}]}{z\sqrt{\sigma_x^2 \sigma_y^2 -c_{xy}}}U-V<0\right\}.
\end{eqnarray}
Let $W(z)\sim\mathcal{N}(\mu_w(z),\sigma_w^2(z))$, with
\begin{equation}
\begin{array}{rcl}
\mu_w(z)&\hskip-3pt = &\hskip-3pt  \displaystyle{\frac{[(1-z)\mu_x -z\mu_y]\sigma_x}{z\sqrt{\sigma_x^2\sigma_y^2-c_{xy}}}}\, ,\\
\sigma_w(z)&\hskip-3pt = &\hskip-3pt  \displaystyle{\left[\frac{[(1-z)\sigma_x^2-z c_{xy}]}{z\sqrt{\sigma_x^2 \sigma_y^2 -c_{xy}}}\right]^2+1}\, ,
\end{array}
\end{equation}
then $\mathbb{P}(W(z)\leq 0)=\mathbb{P}(Z\leq z)$. Therefore, the cumulative distribution function of $W(z)$ describes $\mathbb{P}(Z\leq z)\equiv F(z)$.
\begin{equation}\label{cdfW}
\mathbb{P}(W(z)\leq 0)=\int_{-\infty}^0 \frac{1}{\sqrt{2\pi}\sigma_W(z)} e^{-(z-\mu_W(z))^2/(2\sigma_W(z))^2}~dz.
\end{equation}
Equation (\ref{cdfW}) is an analytic formula for $F(z)$. 
The probability density function is $f(z)=F^\prime(z)$. 
To obtain a Gaussian approximation to the pressure density, we evaluate 
\begin{equation}
\frac{1}{|\Delta p|}f\left(\frac{z-P_i}{P_f-P_i}\right)\, .
\end{equation}

\end{document}